\documentclass[aip,amsmath,amssymb,reprint]{revtex4-1}

\usepackage{graphicx}
\usepackage{dcolumn}
\usepackage{bm}
\usepackage{color}
\usepackage[normalem]{ulem}
\usepackage[T1]{fontenc}

\begin{document}

\title[Assessment of hybrid van der Waals density functionals]{Assessment of two
hybrid van der Waals density functionals for covalent and non-covalent binding of molecules}

\author{Kristian Berland}
\email{kristian.berland@smn.uio.no}
\affiliation{ 
Centre for Materials Science and Nanotechnology (SMN), Department of Physics, University of Oslo, 0316 Oslo, Norway
}%

\author{Yang Jiao}
\affiliation{%
Microtechnology and Nanoscience - MC2, Chalmers University of Technology, \\
SE-41296 Gothenburg, Sweden
}
\author{Jung-Hoon Lee}%
\affiliation{%
Molecular Foundry, Lawrence Berkeley National Laboratory, \\
Berkeley, California 94720, USA
}
\affiliation{
Department of Physics, University of California, \\
Berkeley, California 94720-7300, USA
}
\author{Tonatiuh Rangel}%
\affiliation{%
Molecular Foundry, Lawrence Berkeley National Laboratory, \\
Berkeley, California 94720, USA
}
\affiliation{
Department of Physics, University of California, \\
Berkeley, California 94720-7300, USA
}
%
\author{Jeffrey B. Neaton}
\affiliation{%
Molecular Foundry, Lawrence Berkeley National Laboratory, \\
Berkeley, California 94720, USA
}
\affiliation{
Department of Physics, University of California, \\
Berkeley, California 94720-7300, USA
}
\affiliation{
Kavli energy NanoScience Institute at Berkeley, \\
Berkeley, California 94720-7300, USA
}
\author{Per Hyldgaard}
\email{hyldgaar@chalmers.se}
\affiliation{%
Microtechnology and Nanoscience - MC2, Chalmers University of Technology, \\
SE-41296 Gothenburg, Sweden
}%

\date{\today}

\begin{abstract}
Two hybrid van der Waals density functionals (vdW-DFs) are constructed
using 25\%\, Fock exchange with i) the consistent-exchange 
vdW-DF-cx functional [PRB {\bf 89}, 035412 (2014)] and ii) with the vdW-DF2 functional 
[PRB {\bf 82}, 081101 (2010)]. The ability to describe covalent and 
non-covalent binding  properties  of molecules are assessed. 
For properties related to covalent binding,  atomization energies (G2-1 set), 
molecular reaction energies (G2RC set), as well as ionization energies (G21IP set) are 
benchmarked against experimental reference values.  We find that hybrid-vdW-DF-cx 
yields results that are rather similar to those of the standard non-empirical hybrid 
PBE0 [JCP {\bf 110}, 6158 (1996)], with mean average deviations (MAD) of 4.9 and 5.0 
kcal/mol for the G2-1 set, respectively. In this comparison we used wavefunction-based
Hybrid vdW-DF2 follows somewhat different trends, showing on average significantly larger deviations 
from the reference energies, with a MAD of 14.5 kcal/mol for the G2-1 set. 
Non-covalent binding properties of molecules are assessed using the S22 benchmark set of 
non-covalently bonded dimers and the X40 set of dimers of small halogenated molecules, using
wavefunction-based quantum chemistry results for references.  For the S22 set, hybrid-vdW-DF-cx 
performs better than standard vdW-DF-cx for the mostly hydrogen-bonded systems, with MAD dropping from 
0.6 to 0.3 kcal/mol, but worse for purely dispersion-bonded systems, with MAD increasing from 0.2 to 
0.6 kcal/mol. Hybrid-vdW-DF2 offers a slight improvement over standard vdW-DF2. 
Similar trends are found for the X40 set, with hybrid-vdW-DF-cx performing particularly well 
for binding involving the strongly polar hydrogen halides, but poorly for systems with tiny binding 
energies.  Our study of the X40 set reveals both the potential of mixing Fock exchange with vdW-DF, 
but also highlights shortcomings of the hybrids constructed here.  The solid performance of hybrid-vdW-DF-cx for 
covalent-bonded systems, as well as the strengths and issues uncovered for non-covalently bonded systems, makes 
this study a good starting point for developing even more precise hybrid vdW-DFs. 
\end{abstract}

\maketitle

\section{\label{sec:intro}Introduction}

Hybrid functionals are widely used for computing molecular properties because they generally offer more accurate thermochemical and structural properties 
than non-hybrid semilocal functionals in density functional theory (DFT).\cite{Becke93,Stephens94,PBE0,hse03,hse06,yanai04,Goerigk11,beckeperspective} Through 
various procedures, hybrid functionals mix in a fraction of Fock exchange, 
 introducing a Fock term in the exchange part of a Kohn-Sham density functional. 
Many 
semi-empirical or fitted hybrid functionals have been developed, with B3LYP being one of the most popular.\cite{beckeperspective,Becke93,Stephens94,LYP88} 
One of the most successful hybrids, the PBE0 functional,\cite{PBE0} one the other hand
mixes in 25\%  of Fock exchange with the exchange of the PBE,\cite{pebuer96,BurkePerspective}
 which is based on arguments from many-particle theory.\cite{mixing_rationale,Burke97,Ernzerhof97}

In many molecular complexes, London dispersion forces contribute significantly to the binding energy of the system.  To accurately describe all binding properties of such system, it is desirable to include both non-local correlation and a Fock-exchange fraction within the same functional.\cite{beckeperspective} 
Non-local correlations 
are missing in both semi-local functionals in the generalized-gradient 
approximation (GGA) and in standard hybrid functionals, but they are 
included from the onset in the van der Waals density functional (vdW-DF) 
method.\cite{Dion,thonhauser,lee10p081101,behy14,hybesc14,Thonhauser_2015:spin_signature,Berland_2015:van_waals,roso09} vdW-DF has found a wide user base in 
physics and material science.\cite{rev8,bearcoleluscthhy14,Berland_2015:van_waals}  
It is regularly used in the study of adsorption,\cite{chakarova-kack06p146107,berland11p135001,ErhHylLin15,Liu12p245405,carrasco11p026101} layered systems,\cite{rydberg03p126402,londero12p424212,torbjorn14,Walle14,Lindroth16} molecular crystals,\cite{draxl09p125010,berland10p134705,RanPRB16} and more recently solids with covalent and metallic  bonds.\cite{vdwsolids,Thonhauser_2015:spin_signature,Gharaee2016} However, assessments of how well vdW-DFs describe general properties of molecular systems 
are scarce, which  is likely related to the fact that  hybrid vdW-DFs have so far gone unexplored. 

In this paper we define and assess the performance of two simple, unscreened hybrid van der Waals  functionals, termed vdW-DF-cx0 and vdW-DF2-0, for  molecular systems. 
A number of dispersion-corrected (DFT-D) hybrid functionals do already exist.\cite{doublehybrid2007,Chai2008,rev1,grimme3} 
Those descriptions rely on atom-centered pair-potential corrections to account the non-local correlation. In more advanced variants, the coefficients of these  pair potentials can depend on for instance the electron density, as they are in DFT-TS and more recent extensions\cite{ts09,ts-mbd} and in associated hybrids. \cite{Marom11}  Hybrid variants of the Vydrov-Voorhis (VV) nonlocal density functional, \cite{vv10} which borrows several features of vdW-DF, has also been developed.
All of these schemes include one or several adjustable parameters that can be re-tuned when combined with a hybrid functional to achieve good performance for a set of reference systems. Here, we investigate the performance of two hybrid vdW-DFs 
where we deliberately avoid fitting to reference systems. 

The vdW-DF framework is formally well anchored in many-body physics.
\cite{Dion,dionerratum,thonhauser,hybesc14,Berland_2015:van_waals} However, different versions of the functional have been developed,\cite{Dion,cooper10p161104,vdwsolids,hamada14} such as vdW-DF2\cite{lee10p081101}  and vdW-DF-cx,\cite{behy14,bearcoleluscthhy14}   which emphasize different limits of many-body theory.
In this work we define two new vdW-DF hybrids, vdW-DF-cx0 and vdW-DF2-0, modifying vdW-DF-cx and vdW-DF2 to use 25\% of Fock exchange, in analogy with the design of PBE0.\cite{PBE0,mixing_rationale,Burke97,Ernzerhof97}
The 25\% mixing fraction has been rationalized\cite{Becke93,mixing_rationale,Burke97,Ernzerhof97} in terms of a coupling-constant $\lambda$ integration in the adiabatic-connection formula (ACF).\cite{gulu76,lape77} In a mean-value evaluation of the ACF, using a fraction of
Fock exchange takes care of 
the $\lambda \to 0$  limit, where the functional is exclusively represented by exchange. At the $\lambda\to 1$ end the properties of a system are 
dominated by plasmons,\cite{pinesnozieres} suggesting that functionals of the electron-gas tradition, like PBE, provide robust starting points for hybrid constructions, like PBE0.\cite{mixing_rationale,Burke97} However, the same rationale also applies to the consistent-exchange vdW-DF-cx version,  
which rely on vdW-DF's single plasmon-pole description for specifying the exchange and correlation, except in very inhomogeneous regions.\cite{behy14,bearcoleluscthhy14} In fact, vdW-DF, and this version in particular, can  be viewed as a series expansion of the effective response defined by a GGA-type exchange-correlation hole.\cite{hybesc14,Berland_2015:van_waals}

Even if less strictly enforced, vdW-DF2 is also designed with consistency between the exchange and correlation in mind and can thus also be seen as plasmon based.\cite{lee10p081101,bearcoleluscthhy14,hybesc14,Berland_2015:van_waals}
Whereas vdW-DF-cx has a good consistency between its rather soft exchange functional and the internal exchange that sets the strength of the non-local correlation term, 
vdW-DF2 relies on more rapidly increasing exchange enhancement factors in both the exchange and within the non-local correlation.\cite{behy13,schwinger}  
vdW-DF-cx and vdW-DF2 can therefore be
viewed as two different limits of viable vdW-inclusive density functionals. Thus by benchmarking these two functional for molecular systems, we gain insight into how to best construct an accurate hybrid vdW-DF.

\section{\label{sec:comp_details} computational details}

vdW-DF-cx0 and vdW-DF2-0 were implemented within version 6.0 of the \textsc{quantum espresso} package.\cite{QE} This version includes an implementation of spin-vdW-DF,\cite{Thonhauser_2015:spin_signature}
which is needed to correctly compute atomization energies. 
In all calculations, Martins-Troullier\cite{Troullier91p1993} norm-conserving PBE pseudopotentials of the FHI-\textsc{abinit}\cite{Fuchs1999} kind are employed. The energy cutoff is set to 80 Ry.
Atomic coordinates were relaxed (to 0.0025 eV/{\AA}) in all studies,
except for the S22 set.\cite{s22}  Like in several earlier studies,\cite{lee10p081101,behy14} the internal molecular coordinates were kept fixed for the S22 set and the optimal separations were determined by calculating potential energy curves. 

\section{Results}

\subsection{\label{sec:cov} Covalent binding properties} 

\begin{figure}
  \centering
\includegraphics[width=0.99\linewidth]{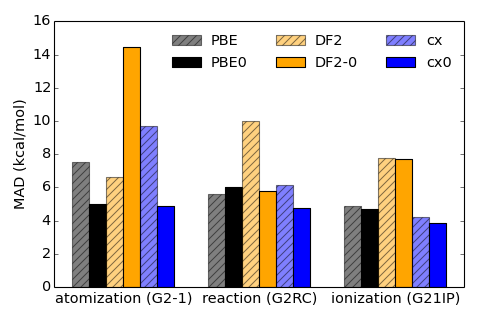}
\caption{Mean absolute deviations of the three benchmark sets considered for covalent molecular  properties.}
  \label{fig:covalent}
\end{figure}

A good hybrid vdW-DF should be able to describe both covalent and non-covalent binding accurately. To assess vdW-DF-cx0 and vdW-DF2-0 ability to describe covalent binding properties, we 
consider atomization energies (in the G2-1 subset) of molecules and chemical reaction energies (in the G2RC subset) of the G2 benchmark set.\cite{G2sets} 
We furthermore benchmark molecular ionization potentials (the G2IP subset).\cite{G2sets} We compare with experimental reference values, in the case of
G2-1 and G2RC subsets back-corrected by wavefunction-based quantum chemistry results for zero-point energies.\cite{G2sets}

Fig.~\ref{fig:covalent} provides an overview of the mean absolute deviation (MAD) of the energies of the three sets, displaying results of vdW-DF-cx0, vdW-DF2-0 and
their non-hybrid equivalents.\footnote{
Results of the G1 set (subset of G2-1) for the non-hybrid spin vdW-DF was presented in the supplementary material of Ref.\ \onlinecite{Thonhauser_2015:spin_signature}. These calculations were based on norm-conserving pseudopotentials rather than ultrasoft pseudopotentials and were compared to the values of the G1\cite{G1set} set rather than the updated reference values of the G2 sets.\cite{G2sets} }
For reference,  PBE and PBE0 results are also shown. The comparison shows that 
among the non-hybrid functionals, PBE performs best, except for atomization energies where vdW-DF2 performs better. 
Among the hybrids, 
vdW-DF-cx0 has the best overall performance, 
slightly better than that of PBE0.
Both vdW-DF-cx0 and PBE0 perform better than the corresponding non-hybrid functionals. In contrast, 
vdW-DF2-0 does not necessarily perform better than vdW-DF2. To gain insight into these trends, we next consider the results of the three sets in more detail.

\begin{table}[h]
\centering
\caption{Benchmark of the 
 atomization energies in the G2-1 set.\cite{G2sets}
The table lists mean relative deviation (MRD), mean absolute relative deviation (MARD), mean deviation (MD), and MAD values of the functionals considered here relative to the experimental references.\label{tab:G21}} 
\begin{tabular}{lrrrrrr}
\hline
\hline
Molecule &PBE & PBE0  &  DF2 & DF2-0 & DF-cx & DF-cx0 \\ 
\hline 
\hline 
MRD [\%]       & 3.63  &  -2.60 &  -0.82 &  -7.25  & 5.82 & -1.40 \\
MARD [\%]      & 5.93  &   4.12 &   4.57 &   9.48  & 7.36 &  3.68 \\ 
MD [kcal/mol]  & 5.82  &  -3.36 &  -2.34 & -12.15  & 9.09 & -1.48 \\
MAD [kcal/mol] & 7.52  &   4.99 &   6.60 &  14.47  & 9.73 &  4.90 \\
\hline 
\hline 
\end{tabular} 
\end{table} 

Table~\ref{tab:G21} gives an overview of our results for atomization energies, in terms of MAD,
mean absolute relative deviation (MARD),  mean deviation (MD), and mean relative deviation (MRD). 
 vdW-DF2-0 is significantly 
less accurate than vdW-DF2, but vdW-DF-cx0 is significantly more accurate than vdW-DF-cx. 
Comparing the MRD values of a non-hybrid with a corresponding hybrid indicates that the atomization energies are on average about 10\,\% lower for the hybrid than for the non-hybrids. 
Since vdW-DF2  underestimates atomization energies on average by about 2\,\%, vdW-DF2-0 underestimates them by about 12\,\%. This systematic underestimation leads to large MAD and MARD values of vdW-DF2-0.
vdW-DF-cx overestimates atomization energies by 9\% on average,  giving it the worst performance among the non-hybrid functionals. However, this overestimation is a good match with the reduction in atomization energies when 
introducing 25\%  Fock exchange, resulting in an MRD value of merely -1.48\%. 
In turn, vdW-DF-cx0 ends up with the smallest MARD value (3.68\,\%) among the tested functionals, somewhat smaller than the PBE0 value 4.12\,\%.

\begin{table}[h]
\centering
\caption{Comparison of the performance of PBE0 and the hybrid van der Waals density functionals, vdW-DF2-0 and vdW-DF-cx0 for chemical reaction energies in the G2RC set.\cite{G2sets}\label{tab:G2RC}} 
\begin{tabular}{lrrrrrr}
\hline
\hline
Molecule &PBE & PBE0  &  DF2 & DF2-0 & DF-cx & DF-cx0 \\ 
\hline 
\hline 
MRD [\%]       & -15.8 &    5.3 &  -31.1 &  -11.4  & -26.4 & -8.6 \\
MARD [\%]      &  37.3 &   33.0 &   59.4 &   36.0  &  47.3 &  30.2 \\ 
MD [kcal/mol]  &  0.73 &  -2.75 &   6.81 &   2.72  &  1.92 & -1.09 \\
MAD [kcal/mol] &  5.63 &   6.06 &  10.01 &   5.76  &  6.14 &  4.77 \\
\hline 
\hline 
\end{tabular} 
\end{table} 

For reaction energies, the hybrid vdW-DFs outperform the non-hybrids. While Table \ref{tab:G2RC} shows that the MAD becomes somewhat smaller, 
the relative deviations are reduced significantly. For instance, the MARD value of 47\% for vdW-DF-cx   drops to 
30\% for vdW-DF-cx0. For vdW-DF2 and vdW-DF2-0, the corresponding numbers are 59\% and 36\%. 
Despite having poorer atomization energies, as discussed earlier, vdW-DF2-0 predicts reaction energies more accurately than vdW-DF2. This suggests a partial error cancellation between the atomization energies of both product and reactant.

\begin{figure}
  \centering
\includegraphics[width=0.99\linewidth]{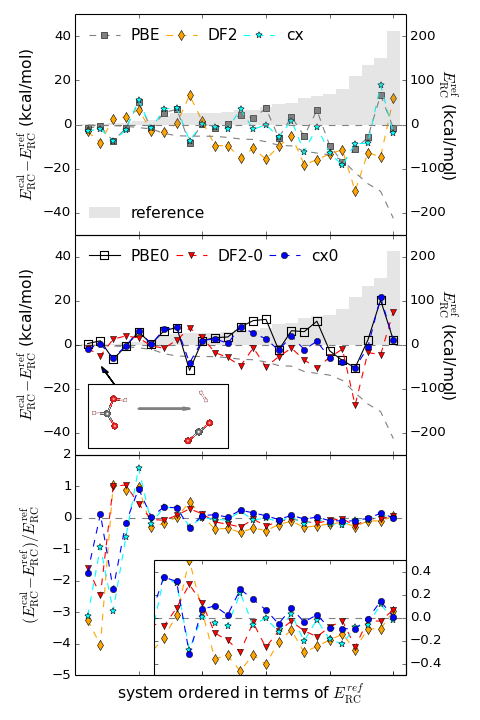}
\caption{
Reaction energies of the G2RC subset of the G2 sets\cite{G2sets} for the different functionals compared here, as ordered by reference reaction energies. These energies are defined as the energy difference between reactants and products, and are indicated by the gray bars in the background (right vertical axis), on a scale five times larger than the deviations. The dashed gray line is eye guide for the magnitude of the reaction energy.
The lowest panel shows the corresponding relative deviations, where the inset provides a smaller span on the vertical axis, for the systems with reference value larger than 20 kcal/mol.}
\label{fig:reactiondetails}
\end{figure}

Fig. \ref{fig:reactiondetails} shows that the deviations grow quite slowly with the size of the reaction energies. 
In the figure, the reactions are ordered according to the size of their respective  reaction energies (gray bars in the upper and middle panels, magnitude given by the right vertical axis). The comparison
shows that the deviations of vdW-DF-cx0 are typically quite similar to those of PBE0, while overall being slightly more accurate. vdW-DF-cx also exhibits deviations that are fairly similar to those of PBE (as shown in panel 1 of Figure \ref{fig:reactiondetails}). In contrast, the deviations of 
vdW-DF2 and vdW-DF2-0 do not follow the same general trends as PBE, PBE0, vdW-DF-cx, and vdW-DF-cx0. 

The bottom panel of figure \ref{fig:reactiondetails} shows that the relative deviations of the different reaction energies. The relative deviations are typically less then 20\% for the reactions with large reaction energies, as shown in the inset of the bottom panel, but they can be much larger for the cases with small reaction energies. In the worst cases, even the sign can be wrong, corresponding to relative deviations less than -1. 
The hybrid variants generally reduce the deviations of the reaction energies. 
As an example, the  energy of the reaction $\rm{CH}_2\rm{O}_2$$\rightarrow$$\rm{CO}_2$+$\rm{H}_2$ (shown in the inset) is  -2.0 kcal/mol.
vdW-DF2 incorrectly predicts this to be an endothermic reaction with reaction energy of 6.1 kcal/mol which drops to 2.9 kcal/mol
when going to vdW-DF2-0. vdW-DF-cx correctly predicts this reaction to be exothermic, but only barely so,  with a reaction energy of -0.1 kcal/mol, vdW-DF-cx0 however gives a value of -2.3 kcal/mol which is only a $12\%$ overestimation of the reference energy. 
The improvements for the systems with small reaction energies are the main reason why the MARD value are significantly  lower for the hybrids, while the reduction in MAD is more modest. 

\begin{table} 
\centering
\caption{Comparison of performance of PBE0 and the hybrid van der Waals density functionals vdW-DF2-0 and vdW-DF-cx0 \label{tab:G21IP} for ionization potential values in the G21IP set.\cite{G2sets}
} 
\begin{tabular}{lrrrrrr}
\hline
\hline
Molecule &PBE & PBE0  &  DF2 & DF2-0 & DF-cx & DF-cx0 \\ 
\hline 
\hline 
MRD [\%]       & -0.35 &   -0.17 &   3.11 &  3.04  & -0.62 &  0.17 \\
MARD [\%]      &  1.98 &    1.97 &   3.28 &  3.30  &  1.72 &  1.67 \\ 
MD [kcal/mol]  & -1.30 &   -0.91 &   7.38 &  7.05  & -1.81 &  0.08 \\
MAD [kcal/mol] &  4.87 &    4.68 &   7.80 &  7.70  &  4.23 &  3.86 \\
\hline 
\hline 
\end{tabular} 
\end{table} 

For ionization potentials, the comparison for the G21IP set, summarized in 
table \ref{tab:G21IP} and Fig.\ \ref{fig:covalent}, shows both the hybrids perform similar or slightly better. vdW-DF-cx and vdW-DF-cx0 perform better than vdW-DF2 and vdW-DF2-0.


The performance of vdW-DF-cx0 for covalent molecular binding properties makes 
this  a promising vdW-DF hybrid functional candidate. vdW-DF2-0 is not as promising, clearly illustrated by its atomization energies.  In a  separate test, we find that the difference between the trends of vdW-DF2-0 and vdW-DF-cx0 for small-molecule atomization energies is predominantly due to the exchange component. For covalent interactions, the performance is rather insensitive to whether vdW-DF1 or vdW-DF2 correlation is employed for a fixed exchange functional. This is in contrast to the case for non-covalent interactions.\cite{behy13} 


\subsection{\label{sec:ncov} Non-covalent molecular binding} 

\begin{figure}
  \centering
\includegraphics[width=0.99\linewidth]{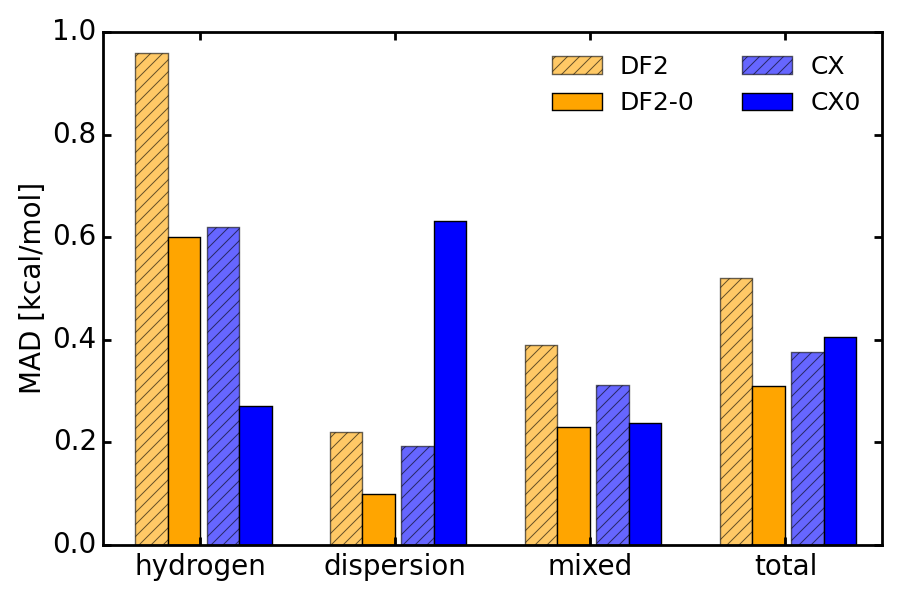}
\caption{The performance of the functionals considered here for the S22 set for inter-molecular binding.}
  \label{fig:noncov}
\end{figure}

Figure~\ref{fig:noncov} shows the results for the S22 set\cite{s22} following the standard division into  hybrid, dispersion, and mixed systems. 
vdW-DF2-0 has smaller MAD than vdW-DF2 for all the three cases. 
vdW-DF-cx0 on other hand has overall a somewhat larger MAD than vdW-DF-cx, increasing from 0.38 kcal/mol to 0.41 kcal/mol. 
Most striking, however, is the significantly worsened performance of vdW-DF-cx0 compared to vdW-DF-cx for purely dispersion-bonded systems 
 --- with MAD increasing from 0.19 to 0.63 kcal/mol ---combined with a much improved performance for the hydrogen bonded systems,  MAD decreasing from 0.62 to 0.27 kcal/mol.
We also find that these poor results for dispersion bonded systems come from a significant overestimation of binding energies. 

The improved performance for hydrogen bonded systems is linked to the fraction of  Fock exchange, as the performance also improves corresponding for vdW-DF2 and PBE. 
In the latter case, MAD shrinks from 1.24 kcal/mol to 1.08 kcal/mol when going from PBE to PBE0. Nonetheless, these numbers are inferior to the vdW-DFs as dispersion forces generally contribute to 
the binding in all kinds of non-covalently bonded dimers, including systems that are mostly hydrogen bonded. We therefore omit PBE and PBE0 in further analysis of non-covalently bounded systems. 

The clear overestimation of vdW-DF-cx0 for dispersion bounded systems shows that Fock exchange and vdW-DF-cx exchange behave quite differently for dispersion-bonded systems, as vdW-DF-cx itself has a rather low MAD. Thus the vdW-DF1 correlation 
(which is also used in vdW-DF-cx) may not be the optimal one for the exchange description used in vdW-DF-cx0. The low vdW-DF-cx0 
MAD for hydrogen-bonded systems nonetheless indicates the promise of mixing a soft 
exchange like that of vdW-DF-cx with Fock exchange.  

The mixed vdW-DF-cx0 performance for the S22 set motivates us to also consider 
the X40 set\cite{X40} of non-covalently bounded dimers involving at least one halogenated molecule.  This set of dimers involve small molecules like methane, and its halogenated derivatives, 
an example being the methane-Cl2 dimer (reference binding energy of 0.69 kcal/mol) or bromobenzenezene-acetone dimer (2.48 kcal/mol. 
They also include the strongly polar hydrogen halides, such as the HF-methanol dimer (5.8 kcal/mol). 

\begin{figure*}[t]
  \centering
\includegraphics[width=0.99\linewidth]{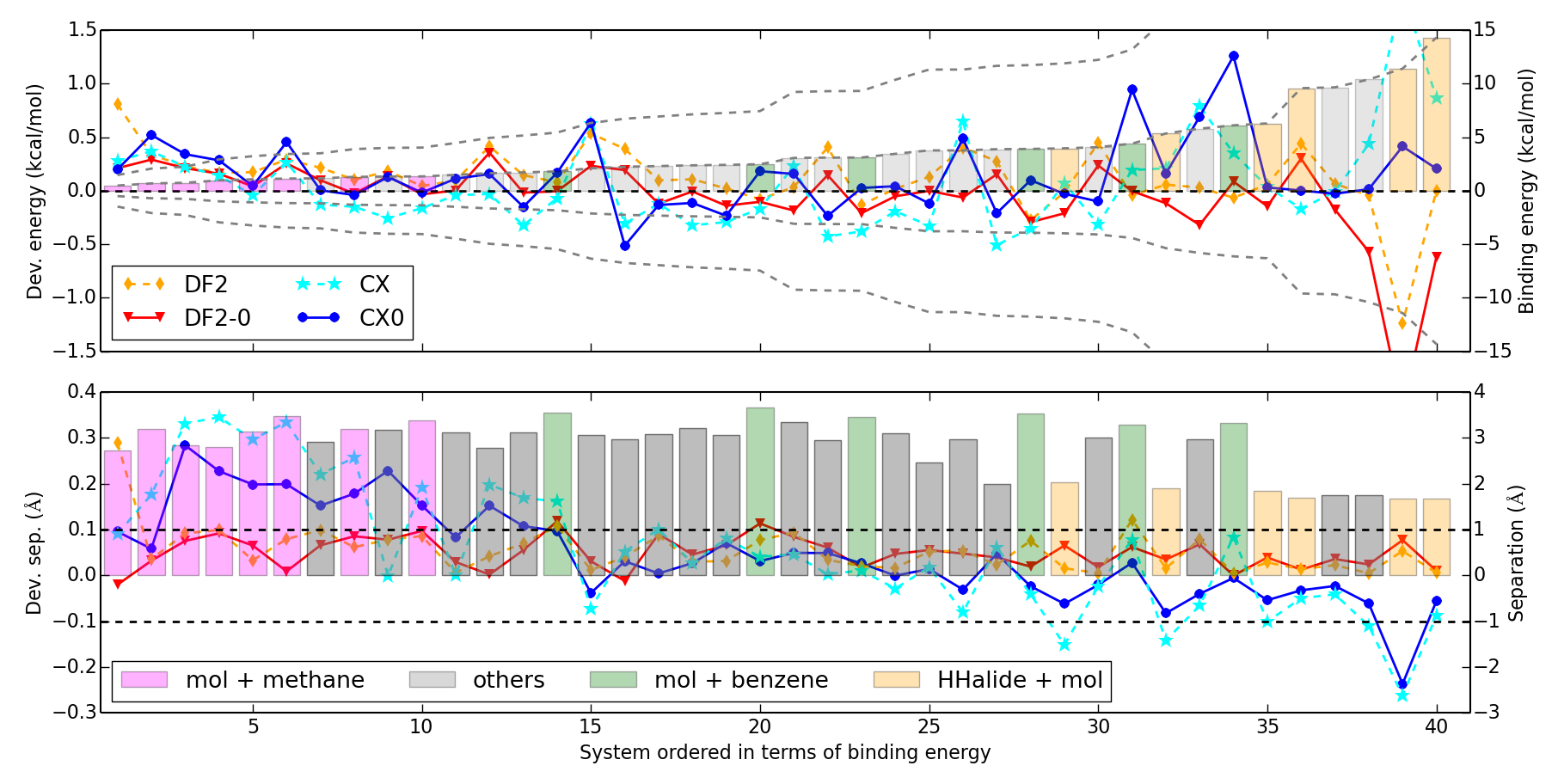}
\caption{Our benchmark of the X40 set of molecule dimers involving halogenated molecules ordered in terms of the reference energies. 
The connected markers in the upper panel indicate the deviation in binding energy of the various molecular dimers. 
The bars show the reference energy on a ten times larger scale (right vertical axis). 
For clarity, the dimers have been ordered into systems involving methane (pink bars), involving benzene (green bars), involving hydrogen halides (orange), and the remainders (gray). The dashed curves correspond to +/- respectively 10\% and 30\% of the binding energy. 
The lower panel shows the intermolecular separations, defined in terms of the shortest atomic separations between dimers. for for each of the systems. 
The bars indicate the full reference separations on a ten times larger scale (right vertical axis).
\label{fig:X40}
}
\end{figure*}

Figure~\ref{fig:X40} shows the deviation in binding energies (upper panel) and binding separation (lower panel) ordered by the magnitude of the reference energy. The bars in upper panel indicate the magnitude of the reference energies, with scale on the right vertical axis. The bar color indicates a sorting of the dimers into molecules  involve methane (pink), benzene (green), hydrogen halide
(orange), or none of these molecules (gray). We organize the X40 set into these subsystems for the following reasons:
For all the dimers of methane and a "halogenated molecule", the binding energies are significantly overestimated and thus dominates the trends in the MRD. 
For benzene-"halogenated molecule" dimers, the aromatic $\pi$-orbitals could contribute to the binding, giving them a somewhat different binding nature. 
For the dimers involving hydrogen halides, the binding is dominated by the strong polarity of the halide.  
Table~\ref{tab:X40} summarizes the trends for the X40 set and these subsystems. 

\begin{table}[h]
\centering
\caption{Summary of our benchmark of the X40 set binding energies. \label{tab:X40}} 
\begin{tabular}{lrrrr}
\hline
\hline
 & DF2 & DF2-0 & DF-cx & DF-cx0 \\ 
\hline 
\hline 
 {\it involving methane} \\
\hspace{0.3cm} MAD[kcal/mol]& 0.27  & 0.15 &  0.20 & 0.24 \\
\hspace{0.3cm} MADR[\%] & 39.3   & 20.1 &  25.7 & 29.9 \\
{\it involving benzene} \\
\hspace{0.3cm}MAD[kcal/mol]& 0.11  & 0.12 &  0.25 & 0.45 \\
\hspace{0.3cm}MADR[\%] & 3.5   & 3.3 &  7.0 & 10.3 \\
{\it involving hydrogen halides} \\
\hspace{0.3cm}MAD[kcal/mol]& 0.30  & 0.57 &  0.54 & 0.14 \\
\hspace{0.3cm}MADR[\%] & 3.0   & 5.9 &  5.1 & 1.6 \\
{\it other dimers} \\
\hspace{0.3cm}MAD[kcal/mol]& 0.24  & 0.17 &  0.28 & 0.15 \\
\hspace{0.3cm}MADR[\%] & 10.3   & 6.0 &  10.4 & 6.2 \\
\hline
{\it Full set} \\
\hspace{0.3cm} MAD[kcal/mol]& 0.22  & 0.21 &  0.32 & 0.24 \\
\hspace{0.3cm} MADR[\%] & 13.2   & 8.1 &  12.4 & 11.7\\
 \hline 
\hline 
\end{tabular} 
\end{table}

For the dimers involving methane, all  the functionals overestimate binding energies, with overall somewhat better performance for the hybrids. The methane molecules in the X40 set are paired with another small molecule, either a diatomic halogen  going from F2 to I2 (system 1, 5, 8 and 10), or halogenated methane derivatives (system 2, 3, 4, and 6). 
Related systems with low  binding energies, such as the chloromethane-formaldehyde dimer (system 7), are also significantly overestimated. 
 The overestimation is evident both in the table and by cases in which the data is outside either the  first or second dashed curves indicating 10\% and 30\% deviation in the upper panel of Fig.\ref{fig:X40}.
 vdW-DF-cx and vdW-DF-cx0 also significantly overestimate binding separations of the methane systems, on average by respectively 0.25\,\AA\, and 0.17\,\AA, as with  
  vdW-DF-cx for the methane dimer of the S22 set.\cite{behy14} Issues for the smallest molecules like methane is a known shortcoming of vdW-DF-cx, which is consequence of using the Langreth-Vosko exchange\cite{lavo87} form in a wide regime of reduced gradients $s$, making it overly repulsive for systems dominated by large $s$ values.\cite{behy14} 
The separations of vdW-DF2 and vdW-DF2-0 are
rather similar for all the systems and both are on average overestimated by about 0.05 Å. This result can be traced to the fact that for molecular dimers, PW86r exchange and Hartree-Fock exchange give quite similar binding curves.\cite{mulela09}

For the dimers involving benzene, vdW-DF2 and vdW-DF2-0 show excellent performance. vdW-DF-cx0 exhibits poorest performance, which arise from a poor description of the  trifluorobenzenezene-benzene dimer and the hexafluorobenzenezene-benzene dimer. For the dimers involving hydrogen halides, which bind strongly, vdW-DF-cx0 exhibits a MADR of only 1.6\%. Even if the binding separations are a bit underestimated, vdW-DF-cx0 is the only functional that does not result in a large deviation of the binding energy of the HCl-methylamine dimer. This is a case for which the binding energy shrinks once Foch exchange is mixed in. 

Considering the remaining 20 dimers, both hybrids exhibit a clear improvement over the non-hybrid functionals with MARD dropping by about 40\% for both vdW-DF-cx0 and vdW-DF2-0.
These dimers involve a range of small organic and halogenated dimers and with varying binding energy and polarity.  For the larger molecules, the overestimation of separations by both vdW-DF-cx and vdW-DF-cx0 turns into an underestimation in energies.  vdW-DF-cx0 typically predicts a similar but somewhat more accurate separations than vdW-DF-cx. 

Evaluating the entire X40 set, we find that hybrid variants do perform better than the non-hybrid functionals. Comparing the MAD and MARD values and Fig.~\ref{fig:X40} shows that vdW-DF2-0 mostly improve the performance  over vdW-DF2 for the systems with small binding energy. 
vdW-DF-cx0 can worsen the binding energies for the systems with small
binding energies, which is also linked to the poorer account of dispersion
bounded systems for the S22 set. It can also improve the performance for
systems with strong polar binding, which is linked to the good performance
for hydrogen-bonded systems of the S22 set.

\section{\label{sec:conclusion} Discussion and outlook}

We have tested the  performance of two hybrid vdW-DFs, assessing both covalent and non-covalent molecular binding properties. 
For covalent molecular binding properties, we find that vdW-DF-cx0 performs very similar to, and even slightly better than PBE0, whereas vdW-DF2-0 is overall less accurate.
This result indicates that the exchange of vdW-DF-cx0 is here a good match for the correlation component.  In turn, this indicates that vdW-DF-cx can indeed, for these properties,  
serve the same role as PBE in mean-value evaluation of the adiabatic connection formula.  Using hybrid functionals also improves the non-covalent interactions for systems 
where hydrogen bonding and other polar effects contribute significantly to the binding.  However, vdW-DF-cx0 overestimates binding energies of purely dispersion-bonded systems 
of the S22 set. 

Insight into a possible mechanism behind the finite vdW-DF-cx0 deviations in describing binding energies of dispersion bonded systems 
can be obtained by recalling the construction of vdW-DF-cx.\cite{behy14} The vdW-DF-cx version rests on a consistent-exchange argument whereby exchange
and correlation are brought in balance by effectively using the same plasmon-pole propagator to define both terms.\cite{bearcoleluscthhy14,hybesc14}
However, in going from vdW-DF-cx to vdW-DF-cx0, we only replaced the exchange contributing to the total energy, but did not update 
the internal exchange within the vdW-DF-cx correlation kernel and to some extent we thereby broke the internal consistency of the 
hybrid functional. Purely dispersion-bonded systems are precisely the regime where the parameterization of the internal functional is 
crucial for the total binding energies of the systems.\cite{behy13} Further progress in constructing more accurate hybrid  vdW-DF versions 
may therefore be achieved by ameliorating this broken consistency. Such an investigation is beyond the scope of the present paper.

Our study shows that hybrid vdW-DFs are promising general-purpose functionals for describing binding in molecules.
vdW-DF-cx0 stands out as an excellent functional for describing covalent and hydrogen/halogen- bonding of molecules, 
but the simple mixing in of 25\% Fock exchange does not preserve the good performance of vdW-DF-cx for typical dispersion bonded dimers nor does it resolve issues of vdW-DF-cx for binding between the smallest molecules. 
This study should therefore motivate further development of hybrid vdW functionals as well as tests of other classes of materials, both dense traditional materials, and sparse vdW materials. In particular  materials where interactions compete and both the sparse- and the dense-matter properties must  be described accurately at the same time\cite{langrethjpcm2009,bearcoleluscthhy14,Berland_2015:van_waals} 
are well suited testing grounds for hybrid vdW-DFs. 
An additional benefit of hybrid vdW-DFs not explored here is the potential for accurate computations of excited-state properties. It is practical and conceptual advantage to be able to compute the excited state properties of vdW bonded materials using the same hybrid vdW-DF version that was first used to relax the 
atomic coordinates.

\section*{\label{sec:ack} Acknowledgment}

The authors thank Zhenfei Liu for insightful discussions.
Work in Sweden supported by the Swedish Research Council (VR), Chalmers 
Area-of-Advance Materials, and the Chalmers e-Science centre. KB acknowledges 
support from the Research Council of Norway (no. 250346). Work by JHL, TR, 
and JBN is supported by the Center for Gas Separations Relevant to Clean 
Energy Technologies, an Energy Frontier Research Center, funded by the 
U.S. Department of Energy, Office of Science, Office of Basic Energy
Sciences, under Award DE-SC0001015. Portions of this work were performed 
at the Molecular Foundry, supported by the Office of Science, Office 
of Basic Energy Sciences, U.S. Department of Energy, under Contract 
DE-AC02-05CH11231.  Computational resources were provided by DOE (LBNL 
Lawrencium and NERSC) and by the Swedish National Infrastructure for 
Computing (SNIC), under contract 2016-10-12, from the 
Chalmers Centre for computing, Science and Engineering (C3SE).

\bibliography{references,new_references}

\end{document}